\documentclass[runningheads]{llncs}
\usepackage[T1]{fontenc}
\usepackage{hyperref}
\usepackage{graphicx}
\usepackage{array}
\usepackage{physics}
\usepackage{caption}
\usepackage{colortbl} 
\usepackage{xcolor}
\begin{document}
\title{
$\rm{IterMask^2}$: Iterative Unsupervised Anomaly Segmentation via Spatial and Frequency Masking for Brain Lesions in MRI}
\titlerunning{$\rm{IterMask^2}$: Iterative Unsupervised Anomaly Segmentation of Brain Lesions}
%
\author{Ziyun Liang$^{1}$, Xiaoqing Guo$^{1}$, J. Alison Noble$^{1}$, Konstantinos Kamnitsas$^{1,2,3}$}
%
\authorrunning{}
%
\institute{$^1$Department of Engineering Science, University of Oxford, Oxford, UK
\\
$^2$Department of Computing, Imperial College London, London, UK \\
$^3$School of Computer Science, University of Birmingham, Birmingham, UK\\
\email{ziyun.liang@eng.ox.ac.uk}
}


\maketitle              
\begin{abstract}
Unsupervised anomaly segmentation approaches to pathology segmentation train a model on images of healthy subjects, that they define as the `normal' data distribution. At inference, they aim to segment any pathologies in new images as `anomalies', as they exhibit patterns that deviate from those in `normal' training data.
Prevailing methods follow the `corrupt-and-reconstruct' paradigm. They intentionally corrupt an input image, reconstruct it to follow the learned `normal' distribution, and subsequently segment anomalies based on reconstruction error. Corrupting an input image, however, inevitably leads to suboptimal reconstruction even of normal regions, causing false positives. 
To alleviate this, we propose a novel iterative spatial mask-refining strategy $\rm{IterMask^2}$. 
We iteratively mask areas of the image, reconstruct them, and update the mask based on reconstruction error. This iterative process progressively adds information about areas that are confidently normal as per the model. The increasing content guides reconstruction of nearby masked areas, improving reconstruction of normal tissue under these areas, reducing false positives. 
We also use high-frequency image content as an auxiliary input to provide additional structural information for masked areas. This further improves reconstruction error of normal in comparison to anomalous areas, facilitating segmentation of the latter. 
\
We conduct experiments on several brain lesion datasets and demonstrate effectiveness of our method. Code is available at: https://github.com/ZiyunLiang/IterMask2

\end{abstract}
\section{Introduction}
\label{sec:intro}

Anomaly segmentation aims to identify patterns that deviate from a `normal' distribution, defined as the distribution of training data. Such algorithms can be used for unsupervised pathology segmentation, where `normal' (training) data are images of healthy tissue, while `anomaly' refers to any pathology. These segmentation methods are termed unsupervised because they do not require any `anomalous' images with pathology nor their manual segmentations for training - they are trained using solely 'normal' images (without pathologies). They can then be used to segment \emph{any} anomalies (pathologies) in new data.
This offers an alternative to the predominant supervised learning paradigm. In the context of brain lesion segmentation, the focus of this study, the common approach is supervised training of a model using manual segmentations for a \emph{specific} pathology \cite{kamnitsas2017efficient,isensee2021nnu,havaei2017brain,stollenga2015parallel}. These models cannot segment other types of pathology. Moreover, it is challenging to train supervised models for rare diseases due to the scarcity of training data, and they are impractical for workflows where the type of pathology that may be in a scan is unknown beforehand. Unsupervised anomaly segmentation can instead enable development of models that can identify \emph{any} pathology and enable applications such as automatic screening for incidental findings \cite{rowley2023incidental}.

\noindent \textbf{Related Work:}
The most prevalent unsupervised anomaly segmentation methods are reconstruction-based. During training, some distortion is applied to the input image and the model learns the `normal' data distribution by learning to reconstruct the image. During inference, anomalies that are not seen during training are challenging to reconstruct, resulting in high reconstruction errors that are used to segment anomalies. For precise anomaly segmentation, \textit{reconstruction errors need to be high on anomalous and low on normal areas.}

Reconstruction-based methods that leverage AutoEncoder (AE) \cite{atlason2019unsupervised,baur2021autoencoders} or Variational AutoEncoder (VAE) \cite{baur2021autoencoders,pawlowski_unsupervised_2018,shen_unsupervised_2019} use compression as distortion and train a model to reconstruct the image.
Another type of reconstruction-based method follows the `corrupt and reconstruct' paradigm: the input is corrupted during \emph{inference}, usually by adding a type of noise, and the model is tasked to reconstruct it.
Such state-of-the-art methods use diffusion models \cite {pinaya2022fast,bercea2023mask}, which corrupt inputs with Gaussian noise.
In both above approaches, increasing input distortion (compression or noise) amplifies reconstruction error of anomalies, since the model has not learned their appearance during training and fails to reconstruct them. This however poses a \emph{sensitivity-precision trade-off}, because more distortion also increases reconstruction error of normal areas, causing false positives \cite{bercea2023aes}. 
\cite{bercea2023mask} proposes to address this trade-off by first generating a mask that covers all anomalies, and then by iteratively reconstructing the masked area with a diffusion model. This relates to our work that also uses an iterative process, but differs conceptually. In that method, the initial mask, which can be suboptimal, is not refined throughout iterations. Furthermore, their generative model does not have any auxiliary guidance for reconstructing the masked content, which hinders faithful reconstruction of normal tissues beneath the mask. Our method effectively addresses these points as we will demonstrate. 
Recently, cross-modality translation has been proposed as an intermediate step for reconstruction \cite{liang2023modality}. A model is trained for translating normal tissues between MRI modalities. It is assumed that during inference the model will fail to translate anomalies, enabling their segmentation. This approach, however, requires multi-modal data. 
Finally, promising results are achieved by denoising methods \cite {kascenas_denoising_2022}. Unlike previously discussed approaches, the input is corrupted by adding noise only during \emph{training} (not inference) and a model is trained to remove it. During inference, it is assumed anomalies will be perceived as noise and will be removed. For this, noise added during training must model the appearance of anomalies and thus requires prior knowledge of their appearance. This is impractical when developing a model for segmenting `any' pathology as herein.

The sensitivity-precision trade-off associated with input corruption in previous works motivates our study: Can we leverage input corruption to \textbf{amplify anomaly reconstruction error} for its segmentation, while \textbf{avoiding information loss in normal areas} to improve their reconstruction and reduce false positives, \textbf{without prior knowledge} about appearance of anomaly?


\noindent \textbf{Contributions:}
This work introduces $\rm{IterMask^2}$, a novel iterative unsupervised anomaly segmentation method that integrates spatial and frequency masking. 
We distort the input using spatial masking and reconstruct masked areas via a model trained solely on normal data.
We \textbf{iteratively refine the spatial masks}, uncovering areas where low reconstruction error indicates normality with high confidence. This gradually re-introduces information about normal areas to further improve the reconstruction of neighbouring normal areas, reduces false positives, and shrinks the mask towards containing solely the anomaly.
To further guide the reconstruction of masked normal areas, we also propose \textbf{providing structural information in the form of high-frequency image components (low-frequency masking)} as additional input. This facilitates more faithful reconstruction of normal areas, whose appearance is predictable from training, in contrast to anomalies whose structure and appearance are unexpected. 
We conducted extensive experiments on several brain lesion datasets that demonstrate the effectiveness of our method. 

\begin{figure}[t!]
\includegraphics[width=\textwidth]{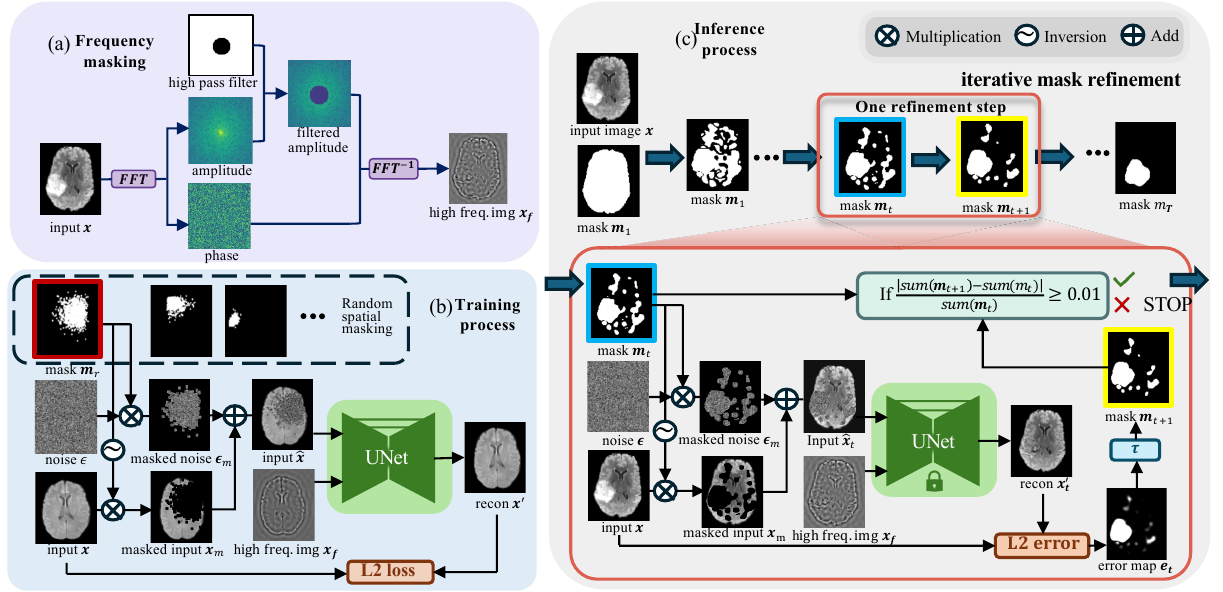}
\caption{Overview of proposed method. (a) The frequency masking strategy that provides structural information for an image via its high frequency components. (b) Training of the reconstruction model. Random spatial masking is applied to input image $\textbf{x}$. High image frequencies from step (a) are given as auxiliary input $\textbf{x}_c$ to the model. (c) shows the iterative mask refinement process during inference. The mask $\textbf{m}_t$ of iteration t gradually shrinks toward the anomaly with the guidance of spatially unmasked area $\textbf{x}_m$ and high image frequencies $\textbf{x}_c$. } \label{fig1}
\end{figure}

\section{Method}
Overview of our method is shown in Fig.~\ref{fig1}. The method follows the `corrupt and reconstruct' paradigm.
To mitigate the sensitivity-precision trade-off of this type of method (Sec.~\ref{sec:intro}), we propose an \textit{iterative spatial mask refinement process} that uses spatial masking as corruption to amplify reconstruction errors on anomalies while minimizing information loss in normal areas by iteratively shrinking a spatial mask toward the anomaly. The mask shrinking process gradually introduces more information about normal-tissue appearance to the model (Fig.~\ref{fig1}(c)).  
In each iteration, the spatially unmasked (ideally normal) area guides the reconstruction of the spatially masked area to shrink towards anomalies. Areas where the error map indicates confidence are considered normal and are removed from the mask for next iterations.
Given the vast diversity of human brain structures, however, reconstructing masked areas without guidance would likely lead to significant reconstruction errors even for normal tissue areas, causing false positives.
Therefore, we further propose extracting structural information from the image and using it as auxiliary input to the network to guide the reconstruction of masked areas. The structural information is \textit{high-frequency image components from the Fourier domain}, as shown in Fig.\ref{fig1}(a). The \textit{training process} is shown in Fig.\ref{fig1}(b). Below we describe these components in detail.

\subsubsection{Iterative Spatial Mask Refinement for Inference}

During inference, the model aims to segment any pathology in a test dataset of brain scans $D_{te}=\left\{\mathbf{x}^{i}\right\}^N_{i=1}$. Assuming that the model was trained on data containing only healthy tissues, which define the `normal' distribution, at inference pathologies are to be segmented as `anomalous' areas with patterns deviating from the normal data distribution. For this, we use model $f$ that reconstructs spatially masked image areas. Given an input image $\mathbf{x}$ that we aim to segment, an initial mask $\textbf{m}_1$ is created that covers the entire brain. We will shrink the mask iteratively, with $\textbf{m}_t$ being the mask at iteration $t$. The masked area is covered with Gaussian noise $\epsilon$, while the unmasked area is copied from the input image. Therefore, the input to the model is $\hat{\mathbf{x}}_t = \mathbf{m}_t \times \epsilon + (1-\mathbf{m}_t) \times \mathbf{x}$. 
The model reconstructs the image giving output $\mathbf{x}_t' = f(\hat{\mathbf{x}}_t, \mathbf{x}_f)$. Here, auxiliary input $\mathbf{x}_f$ is high-frequency image content, described in the next part of this section.
We then compute the error map between $\mathbf{x}_t'$ and original image via $L_2$ distance $\mathbf{e}_t = \lVert \mathbf{x}_t' -\mathbf{x} \rVert _2$. Since we are constantly shrinking the mask towards anomalies, we only consider the error under masked area, $\mathbf{e}_t' = \mathbf{e}_t \times \textbf{m}_t$. 
To determine how to shrink the mask for the next iteration, we compare $\mathbf{e}_t'$ with threshold $\tau$. Pixels below $\tau$ are considered confidently normal as per the model and are removed from the mask.
A detailed analysis on $\tau$ can be found in Section \ref{sec:exp}.
The first iteration is treated as a special case for convenience. Mask $\textbf{m}_2$ for the second iteration is calculated using the 40th percentile of the first iteration's error map $\mathbf{e}_1'$ as threshold $\tau$, unmasking 40\% of the brain as healthy, for faster mask convergence.
The iterative process terminates automatically when the mask shrinks less than 1\%.

\subsubsection{Frequency Masking:} 
\label{method:subsec1}

Frequency domain analysis is fundamental in medical imaging, such as for MRI reconstruction, image denoising and more. Low frequencies of an image's Fourier spectrum capture the mean image intensity (DC signal) and intensities of large image components. High frequencies, on the other hand, capture fine details like edges, boundaries between tissues, and fine outlines of structures \cite{chapelle2019}. 
Amplitude captures low-level statistics while phase is necessary for restructuring the Fourier components back to the original image  \cite {oppenheim1981importance,oppenheim1979phase}. 

Our method aims to reconstruct areas of an image masked in the spatial domain. To guide the reconstruction process of
masked areas and enable more faithful reconstruction of normal tissues, we provide structural image information to the model in the form of high-frequency and phase image components. This reduces false positives by lowering reconstruction error of normal tissue without accompanied decrease for anomalous areas, because the model has not learned how to recreate appearance of anomalies from their high frequency components.

Given a 2D image $\textbf{x}$ (slice of 3D scan) with size $H \times W$, we use the Discrete Fourier Transform $\mathcal{F}$ to map the input image to the frequency domain, obtaining:
$\mathcal{F}(\textbf{x})(a,b)=\sum_{h=0}^{H-1}\sum_{w=0}^{W-1}e^{-i2\pi(\frac{ah}{H}+\frac{bw}{W})}\textbf{x}(h,w)$, where $\textbf{x}(h,w)$ is the spatial pixel value, and the output is complex frequency value. Amplitude and phase are computed by $\mathcal{A} = \left| \mathcal{F}(\textbf{x})(a,b) \right|$ and $\phi=tan^{-1}\frac{Im(\mathcal{F}(\textbf{x})(a,b))}{Re(\mathcal{F}(\textbf{x})(a,b))}$, where $Im$ and $Re$ refers to imaginary and real of the complex value respectively. 
To capture high-level structural information, we use a high-frequency filter to mask out low-frequency amplitude while keeping all the phase information. Our high-pass filter for amplitude masking is:
\begin{equation}
N(u,v) = \left\{
\begin{aligned}
1, & \quad if \quad d((c_u,c_v),(u,v))<r\\
0, & \quad else
\end{aligned}
\right.
,\quad \mathcal{A}_m = \mathcal{A} \times N
\end{equation}
Finally, the masked amplitude and phase are combined by $A_m * e^{j*\phi}$
before inverse Fourier transform $\textbf{x}_f = \mathcal{F}^{-1}(\textbf{x})$ map 
it back to image space.

\subsubsection{Spatial Random Masking for Training:}
To realize the aforementioned process, we train a UNet as model $f$, which takes spatially masked images from the training set $D_{tr}=\left\{\mathbf{x}^{i}\right\}^N_{i=1}$ as input and outputs the reconstructed result, with the high image frequencies $x_f$ as auxiliary input. As there are no anomalies in training, we generate randomly shaped masks for training. For this, we first select a random point $(\mu_x,\mu_y)$ in brain area and generate a multivariate 2D Gaussian distribution with the selected point as center. The probabilistic density function of the Gaussian distribution is $p(x,y)=\frac{1}{2\pi \sigma^2}e^{-[(x-\mu_x)^2+(y-\mu_y)^2]/(2\sigma^2)}$, where $(x,y)$ is the position. We sample the covariance $\sigma$ for generating a Gaussian distribution with a dense region that won't cover the whole brain. Then, the probabilistic density function $p(x,y)$ is used as the probability to sample masked patches from the image. The patches have random side lengths $l={4,8,16}$, and the probability of a patch sampled at position $(x,y)$ is $\frac{p(x,y)}{\sum_{x,y=1:X,Y}p(x,y)}$. After sampling around 1000 patches, random-shaped masks $\mathbf{m}_r$ are generated as shown in Fig.\ref{fig1}(b). For input image $\textbf{x}$, masked input to the model is $\hat{\textbf{x}} = \epsilon \times \mathbf{m}_r + \mathbf{x} \times (1-\mathbf{m}_r)$, where $\epsilon$ is Gaussian noise. The model's output is reconstruction $x'=f(\hat{\textbf{x}}, x_f)$. The training loss is the reconstruction error $L = \lVert x'-\mathbf{x} \rVert _2$. 

Note that we treat as a special case the first iteration of the iterative process described previously in this section, because the whole image $\textbf{x}$ is masked for $t=1$. For this, we train a second UNet separately. This model receives as input only the high-frequency image $\textbf{x}_f$ and is trained to reconstruct whole input $\textbf{x}$. 

\section{Experiment} \label{sec:exp}

\textbf{Setup:}
We evaluate our method by training our model on healthy (i.e. `normal') 2D slices and segment brain pathologies at test time. We use the \textbf{BraTS2021} dataset \cite {bakas_advancing_2017,baid_rsna-asnr-miccai_2021} with 1251 tumor cases, \textbf{ATLAS} v2.0 \cite {liew2022large} with 655 stroke lesion cases, and SSIM of \textbf{ISLES} 2015 \cite{maier2017isles} with 28 stroke lesions. For BraTS, we use the same data pre-processing and selection process as described in \cite{liang2023modality}. For ATLAS and ISLES, we perform skull-strip and apply z-score intensity normalization as pre-processing. 
All lesion labels are merged into a single class. 

\textbf{Results:}
To evaluate segmentation quality, we use the dice coefficient (DSC), sensitivity, and precision. To evaluate the reconstruction performance in `normal' areas, we measure the structural similarity by computing the SSIM score between the original image and the reconstructed image after excluding the anomaly regions. 
We compare our method to a variety of baseline methods, including the autoencoder-based method (AE) \cite{baur2021autoencoders}, diffusion model-based method (DDPM) \cite{pinaya2022fast}, diffusion model with iterative resampling AutoDDPM \cite{bercea2023mask}, modality translation Cyclic-UNet \cite{liang2023modality}, and denoising autoencoder DAE \cite{pinaya2022fast}. In Cyclic-UNet, the middle modalities used are T1 (FLAIR as input) and FLAIR (otherwise).
\begin{table}[t!]
\centering
\caption{Performance for different anomaly segmentation methods on all modalities of BraTS. Grey-shaded cells use the best threshold per-image. Best in bold.} \label{tab1}
\scalebox{0.78}{
\begin{tabular}{l|cccc|cccc|cccc|cccc}
\hline
Modality & \multicolumn{4}{c|}{\textbf{FLAIR}} & \multicolumn{4}{c|}{\textbf{T1CE}} & \multicolumn{4}{c|}{\textbf{T2}} & \multicolumn{4}{c}{\textbf{T1}} \\ \hline
Metrics & DSC & Sens. & Prec. & SSIM & DSC & Sens. & Prec. & SSIM & DSC & Sens. & Prec. & SSIM & DSC & Sens. & Prec. & SSIM \\ \hline
AE\cite{baur2021autoencoders} & 33.4 & 54.0 & 27.0 & 0.941 & 32.3 & 50.1 & 26.4 & 0.957 & 30.2 & 62.8 & 21.1 & 0.937 & 28.5 & \textbf{94.9} & 17.4 & 0.953 \\
DDPM\cite{pinaya2022fast} & 60.7 & 57.8 & 69.8 & 0.980 & 37.9 & 35.2 & 46.2 & 0.984 & 36.4 & 33.6 & 44.2 & 0.984 & 29.4 & 32.1 & 31.5 & 0.977 \\
AutoDDPM\cite{bercea2023mask} & 55.5 & 57.5 & 58.7 & 0.937 & 36.9 & 58.9 & 28.9 & 0.903 & 29.7 & 53.0 & 22.3 & 0.907 & 33.5 & 61.1 & 24.7 & 0.910 \\
Cycl.UNet\cite{liang2023modality} & 65.0 & 63.4 & 73.9 & 0.864 & 42.6 & 47.5 & 42.9 & 0.923 & 49.5 & 48.8 & 53.4 & 0.831 & 37.0 & 45.8 & 35.2 & 0.926 \\
DAE\cite{kascenas_denoising_2022} $\left[ 0,+\infty \right]$ & 79.7 & 79.1 & \textbf{84.5} & 0.926 & 36.7 & 42.0 & 36.2 & 0.844 & 69.6 & 68.1 & 75.3 & 0.903 & 29.5 & 61.2 & 20.5 & 0.860 \\
DAE $\left[ -\infty,0 \right]$ & 28.5 & \textbf{94.9} & 17.3 & 0.875 & 34.7 & 37.9 & 39.3 & 0.966 & 28.5 & \textbf{94.9} & 17.4 & 0.885 & 47.9 & 53.7 & 50.9 & 0.958 \\
DAE$\left[ -\infty,+\infty \right]$ & 73.7 & 72.9 & 80.5 & 0.926 & 46.3 & 47.9 & 51.5 & 0.969 & 60.4 & 58.5 & 69.1 & 0.897 & 44.5 & 48.0 & 47.5 & 0.964\\
$\rm{IterMask^2}$ & \textbf{80.2} & 81.3 & 83.3 & \textbf{0.985} & \textbf{61.7} & \textbf{59.1} & \textbf{70.9} & \textbf{0.997} & \textbf{71.2} & 74.4 & \textbf{72.9} & \textbf{0.994} & \textbf{58.5} & 56.6 & \textbf{67.6} & \textbf{0.996}\\
\hline
\rowcolor[gray]{.93}DAE\cite{kascenas_denoising_2022} $\left[ 0,+\infty \right]$ & 84.2 & 83.1 & 87.0 & 0.926 & 42.1 & 52.5 & 42.8 & 0.844 & 74.5 & 72.1 & 80.1 & 0.903 & 35.3 & 64.3 & 28.3 & 0.860 \\
\rowcolor[gray]{.93}$\rm{IterMask^2}$ & \textbf{85.9} & \textbf{85.7} & \textbf{87.4} & \textbf{0.991} & \textbf{67.7} & \textbf{66.4} & \textbf{76.4} & \textbf{0.997} & \textbf{78.3} & \textbf{75.6} & \textbf{84.0} & \textbf{0.995} & \textbf{64.9} & \textbf{65.2} & \textbf{71.4} & \textbf{0.995}\\
\hline
\end{tabular}
}
\end{table}

We first evaluated performance for segmenting tumors as anomalies across all modalities of the \textbf{BraTS} dataset. As shown in white-shaded parts of Tab.\ref{tab1}, our method exhibits promising performance compared to existing methods that do not include prior information during training (i.e. AE, DDPM, AutoDDPM, Cycl.-UNet), improve dice by 15.2\%, 19.1\%, 21.7\%, 21.5\% in FLAIR, T1CE, T2, and T1 modalities respectively. 
DAE (shown in the $\left[ 0,+\infty \right]$ row shaded white in Tab.\ref{tab1}) demonstrates promising performance in FLAIR and T2 modalities, where tumors are hyper-intense, but fails to outperform other baselines methods in T1CE and T1.
This is due to hyper-intense noise being used as prior during training (i.e. by learning to remove noise of range $\left[ 0,+\infty \right]$).
We further tested DAE with different intensity priors by adjusting the intensity of added noise. 
In our method, without introducing prior intensity information, 
we outperform DAE with all kinds of noise in the dice score. More specifically, we compared our method to DAE when the intensity prior matches with the anomaly ($\left[-\infty, 0 \right]$ range of noise for T1 and $\left[-\infty, +\infty \right]$ for T1CE). We improved dice by 0.5\% and 1.6\% in FLAIR and T2 with hyper-intense noise, and 15.4\% and 10.6\% in T1CE and T2, where the noise intensity is similar to normal tissues.
In terms of reconstruction performance, our method improves SSIM to 0.985 in FLAIR, 0.997 in T1CE, 0.994 in T2, and 0.996 in T1, showing our iterative mask-shrinking approach can reconstruct normal areas better than other existing methods. 
Fig.\ref{fig:fig2} shows the qualitative result of our proposed method compared to baselines. 

\begin{figure}[t!]

  \begin{minipage}[t]{.55\linewidth}
    \centering
     \captionof{table}{Performance on ISLES and ATLAS datasets. Best marked in bold.}\label{tab:tab2}
\scalebox{0.77}{
\begin{tabular}{l|cccc|cccc}
\hline
 Modality & \multicolumn{4}{c|}{\textbf{ISLES FLAIR}} & \multicolumn{4}{c}{\textbf{ATLAS T1}} \\ \hline
Metrics & \multicolumn{1}{c}{DSC} & \multicolumn{1}{c}{Sens.} & \multicolumn{1}{c}{Prec.} & \multicolumn{1}{l|}{SSIM} & \multicolumn{1}{l}{DSC} & \multicolumn{1}{l}{Sens.} & \multicolumn{1}{l}{Prec.} & \multicolumn{1}{l}{SSIM} \\ \hline
AE\cite{baur2021autoencoders} & \multicolumn{1}{c}{21.5} & \multicolumn{1}{c}{0.475} & \multicolumn{1}{c}{0.167} & \multicolumn{1}{c|}{0.791} & \multicolumn{1}{c}{11.9} & \multicolumn{1}{c}{40.7} & \multicolumn{1}{c}{8.6} & \multicolumn{1}{c}{0.954} \\
DDPM\cite{pinaya2022fast} & 44.0 & 50.3 & 43.9 & 0.937 & 20.2 & 22.6 & 26.5 & 0.987 \\
AutoDDPM\cite{bercea2023mask} & 41.1 & 52.6 & 42.9 & 0.862 & 12.7 & \textbf{47.4} & 9.2 & 0.892 \\
Cycl.UNet\cite{liang2023modality} & 53.4 & 50.5 & \textbf{59.4} & 0.842 & - & - & - & - \\
DAE\cite{kascenas_denoising_2022} & 51.8 & 61.9 & 52.5 & 0.914 & 11.1 & 33.4 & 32.7 & 0.995 \\
$\rm{IterMask^2}$  & \textbf{55.1} & \textbf{59.8} & 58.0 & \textbf{0.964} & \textbf{35.3} & 34.3 & \textbf{41.7} & \textbf{0.996} \\ \hline
\rowcolor[gray]{.93}DAE\cite{kascenas_denoising_2022} & 56.5 & \textbf{66.8} & 55.6 & 0.905 & 11.1 & 33.4 & 32.7 & \textbf{0.995} \\
    \rowcolor[gray]{.93}$\rm{IterMask^2}$  & \textbf{59.6} & 69.1 & \textbf{61.7} & \textbf{0.956} & \textbf{47.5} & \textbf{57.7} & \textbf{54.2} & 0.989 \\ \hline
    \end{tabular}}
  \end{minipage}
  \hspace{0.02\textwidth}
  \begin{minipage}[t]{.45\linewidth}
    \centering
    \captionof{table}{Ablation study.
    $DS_{S,M,L}$ represents Dice score of small, medium and large lesions.}\label{tab:tab3}
    \scalebox{0.77}{
 
\begin{tabular}{lcccccc}
\hline
\multicolumn{1}{l|}{Modality} & \multicolumn{6}{c}{\textbf{BraTS FLAIR}} \\ \hline
\multicolumn{1}{l|}{Metrics} & DSC & Sens. & Prec. & $\rm{DS_S}$ & $\rm{DS_M}$ & $\rm{DS_L}$ \\ \hline
\multicolumn{1}{l|}{-freq} & 64.2 & 69.4 & 63.6 & 20.7 & 63.4 & 72.1 \\
\multicolumn{1}{l|}{-spat} & 75.1 & 73.8 & 79.6 &  46.6 & 77.9 & 77.0 \\ \hline
\multicolumn{1}{l|}{DAE\cite{kascenas_denoising_2022}} & 79.7 & 79.1 & \textbf{84.5} & 47.3 & \textbf{81.5}& 83.0 \\
\multicolumn{1}{l|}{$\rm{IterMask^2}$} & \textbf{80.2} & \textbf{81.3} & 83.3 & \textbf{55.6} & 80.8 & \textbf{83.6} \\ \hline
\rowcolor[gray]{.93}\multicolumn{1}{l|}{DAE\cite{kascenas_denoising_2022}} & 84.2 & 83.1 & 87.0 & 60.6 & 85.3 & 87.0 \\
\rowcolor[gray]{.93}\multicolumn{1}{l|}{$\rm{IterMask^2}$} & \textbf{85.9} & \textbf{85.7} & \textbf{87.4} & \textbf{65.7} & \textbf{86.3} & \textbf{88.7} \\ \hline
\end{tabular}
    

    }
  
  \end{minipage}\hfill
\end{figure}

We further evaluated our method on stroke lesion segmentation in the ISLES and ATLAS datasets, as shown in the white shaded parts in Tab.~\ref{tab:tab2}. Our method shows superior performance for almost all the evaluated metrics compared to baselines, demonstrating the generalizability of IterMask$^{2}$ to different anomalies. 


\textbf{Ablation Study:} We assess the effectiveness of the iterative spatial mask refinement and frequency-masking as in Tab.\ref{tab:tab3}. Respectively, we removed frequency guidance after the initial step (\textit{`-freq'} row) and we used high image frequencies as input to generate a single-step prediction (removing the iterative spatial mask refinement process) (\textit{`-spat'} row). 
Dice score drops by 16.0\% and 5.1\% respectively, showing the effectiveness of both components. 

\begin{figure}[t!]
  \begin{minipage}[b]{.63\linewidth}
    \centering
    \includegraphics[width=0.9\linewidth]{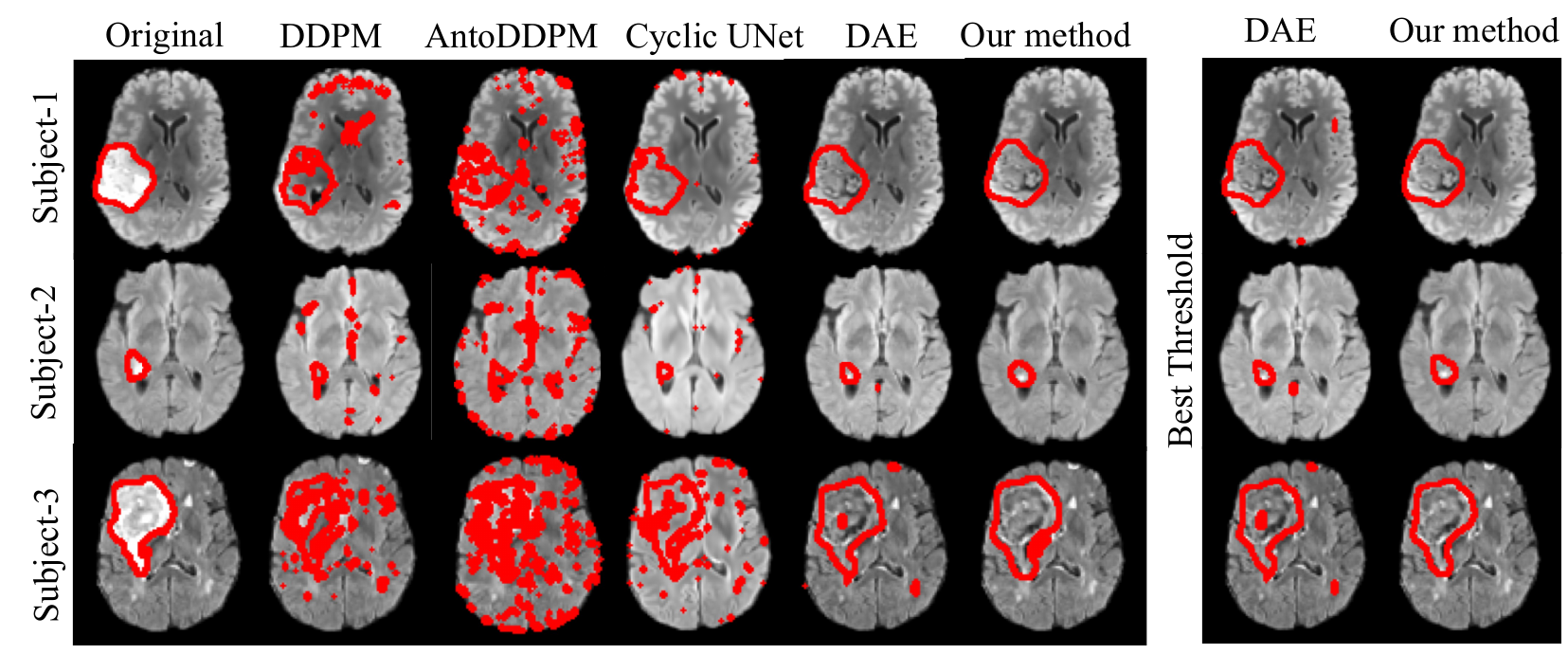}
    \captionof{figure}{Visualization result for FLAIR(BraTS).}
    \label{fig:fig2}
  \end{minipage}
  \hspace{1pt}
  \begin{minipage}[t]{.35\linewidth}
    \centering
    \vspace{-110pt}
    \includegraphics[width=0.98\linewidth]{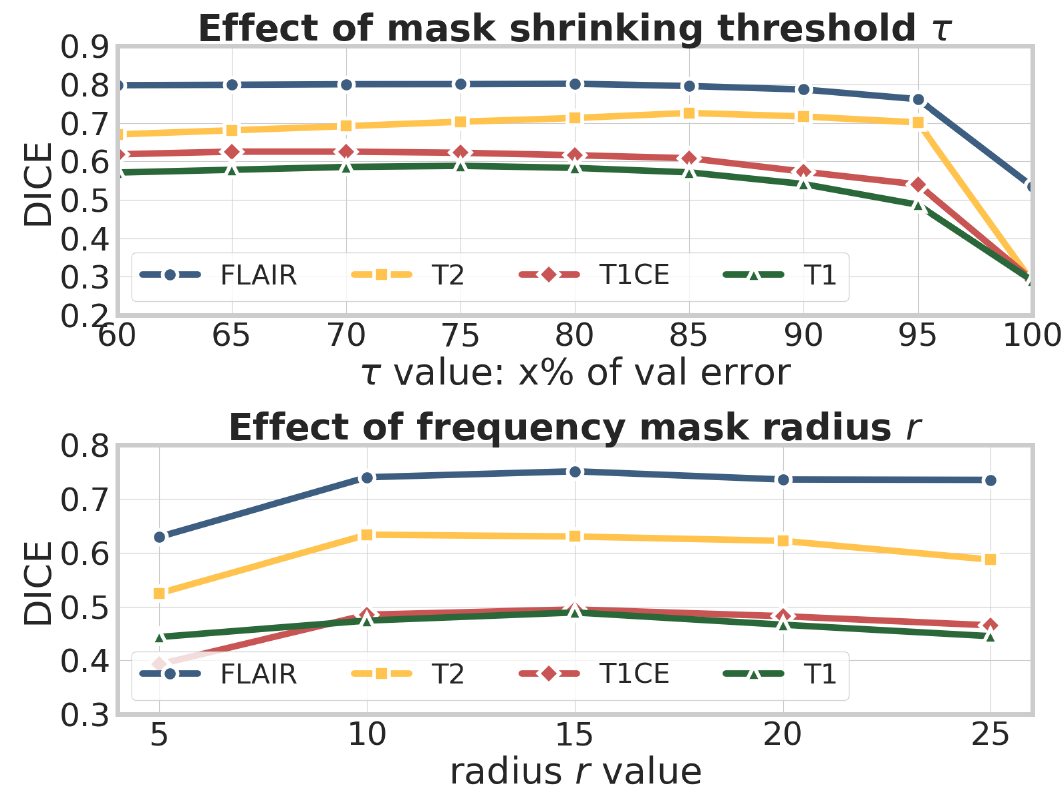}
    \captionof{figure}{Sensitivity Analysis.}
    \label{fig:fig3}
  \end{minipage}\hfill
\end{figure}

\textbf{Sensitivity Analysis:} 
We first analyze the sensitivity of our model to the \textbf{mask shrinking threshold $\tau$} in the iterative spatial mask refinement process. 
We choose $\tau$ from the healthy validation set's error map obtained by applying random masks in the same way as training. 
As shown in Fig.~\ref{fig:fig3}, we experimented with $\tau$ ranging from 60-100 percentiles of the error map's masked area on all modalities of BraTS. 
Observe that our model demonstrates stability across a broad spectrum of $\tau$, showcasing a notable level of fault tolerance. 
This is credited to our way of computing the final segmentation, where we use the last iteration's reconstruction rather than directly using the final mask. So the algorithm has some degree of fault tolerance when the mask is larger than the anomaly. In Tab.\ref{tab1} \ref{tab:tab2} \ref{tab:tab3}, $\tau$ is set at 80 percentile of the error map in BraTS and ISLES, and 70 in ATLAS, values found adequate in preliminary experiments prior to this analysis (thus not over-optimized).
Furthermore, we analyzed sensitivity of our algorithm to the frequency masking \textbf{radius $r$}. 
In the experiment, we set high image frequencies as input to reconstruct images in a single step, showing their influence on the reconstruction process. 
Observe from Fig.\ref{fig:fig3} that the model's performance is stable to radius from 10-25 in all the modalities of BraTS. In all our reported experiments, we set the radius to 15, similar to the setting in \cite{xie2022masked}.

\textbf{Human-AI collaboration:}
The threshold $\tau$ is an adjustable parameter, providing clinicians with the flexibility to interactively adjust the segmentation  
if the result using the optimized threshold gives unsatisfactory results. 
Moreover, our model's mask-shrinking process offers an intuitive interface for human interaction by showing the intermediate decision process. 
Therefore, we tested our model by incrementally raising $\tau$ value during the shrinking process and selected the iteration with the optimal performance per image for evaluation, obtaining improved performance, as shown in gray-shaded cells of Tab.\ref{tab1}~\ref{tab:tab2}~\ref{tab:tab3}. We further compared our models' performance with the best baseline method's optimal performance (DAE), also optimizing the threshold per image when choosing the final segmentation. We found that our model constantly achieves better performance on all datasets, revealing the promising potential to interact with clinicians. 
Furthermore, we categorized the tumor size into small (size<200), medium, and large (size>800), demonstrating human collaboration has the potential to notably improve performance in detecting small anomalies(as shown in gray-shaded units in Tab.\ref{tab:tab3}). When the anomaly is small, the model can prompt clinician intervention, potentially improving prediction accuracy with minimal human effort.


\section{Conclusion}
This paper introduces a novel unsupervised anomaly segmentation algorithm $\rm{IterMask^2}$ for segmentation of brain lesions in MRI. The process uses iterative spatial mask refinement and is guided by structural information from high-frequency image components. 
Extensive experiments on brain tumor and stroke lesion datasets demonstrated the effectiveness of the method in comparison to previous works. Given the very promising results, future work could seek to extend the framework for processing 3D images instead of slices processed herein.

\section{Acknowledgement}
ZL is supported by a scholarship provided by the EPSRC Doctoral Training Partnerships programme [EP/W524311/1]. The authors acknowledge UKRI grant reference [EP/X040186/1] and EPSRC grant [EP/T028572/1].
The authors also acknowledge the use of the University of Oxford Advanced Research Computing (ARC) facility in carrying out this work(http://dx.doi.org/10.5281/zenodo.22558). 

\bibliographystyle{splncs04}
\bibliography{references}




\end{document}